\RequirePackage{lineno}
\documentclass[
  ,draft            
  ]
  {aipproc}
\usepackage{natbib}
\layoutstyle{6x9}
\usepackage{float}
\usepackage{graphicx}
\usepackage{lineno}
\bibpunct{[}{[}{,}{n}{,}{,}

\begin{document}
\title{Neutral Current Elastic Interactions in MiniBooNE}


\author{Ranjan Dharmapalan for the MiniBooNE Collaboration}{
  address={dharm001@crimson.ua.edu}
}

\begin{abstract}
Neutral Current Elastic (NCE) interactions in MiniBooNE are discussed.  In the neutrino mode  MiniBooNE reported: the flux averaged NCE differential cross section as a function of four-momentum transferred squared, an axial mass ($M_A$) measurement, and a measurement of the strange quark spin content of the nucleon, $\Delta s$.  In the antineutrino mode we present the background-subtracted data which is compared with the Monte Carlo predictions.
\end{abstract}

\maketitle


\section{Neutral Current Elastic Interactions and MiniBooNE}
Neutral current elastic (NCE) scattering occurs when a neutrino interacts with a particle by exchanging a $Z^{0}$ boson and the particle remains in its original state.
$$
\nu N \to \nu N
$$
NCE interactions can be used to search for the strange quark spin content of the nucleon as the NCE process is sensitive to nucleon isoscalar weak current. 

The MiniBooNE experiment is a short baseline neutrino oscillation experiment\cite{MB_intro_1,MB_intro_2}
 at Fermilab. At MiniBooNE NCE interactions account for $17\%$ of total events in the $\nu$ mode and around $14\%$ in $\bar\nu$ mode. In MiniBooNE, neutrino interactions are simulated with the NUANCE-v3 event generator \cite{nuance} where a relativistic Fermi gas (RFG) model of Smith and Moniz \cite{Smith_Moniz} is used to characterize the NCE process. Within this model, two parameters are employed: the nucleon axial mass $M_A$  and the NCE scattering strange quark contribution to the axial form factor, $\Delta s$. Nuclear effects can make the axial mass an "effective" value and not the same as scattering from individual nucleons. MiniBooNE can run in both $\nu$ or $\bar\nu$ modes by virtue of a magnetic horn which sign selects charged mesons. 


\section{neutrino mode results}

\begin{figure}
\includegraphics[height=.3\textheight]{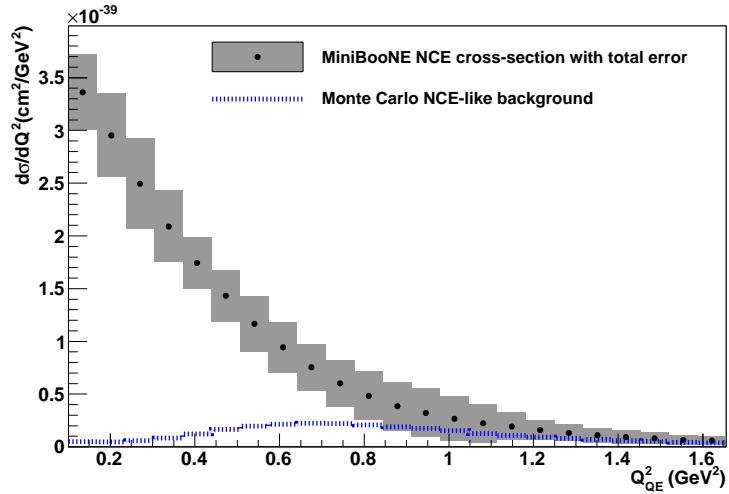}
\caption{\textit{(color online)} The MiniBooNE NCE ($\nu N\rightarrow\nu N$)  flux averaged differential cross section on $CH_{2}$ as a function of $Q^{2}_{QE}=2M_{N}\sum_{i}T_{i}$ where we sum the true kinetic energies of all final state nucleons produced in the NCE interaction. The blue dotted line is the predicted spectrum of NCE-like background which has been subtracted out from the total NCE-like differential cross section. }
\label{fig:1}
\end{figure}
Running in $\nu$ mode MiniBooNE collected $6.46\times 10^{20}$ POT resulting in 94,531 NCE events which passed the NCE selection criteria. This is the largest NCE sample collected to date, with an efficiency of 35$\%$ and  purity of 65$\%$. After background subtraction MiniBooNE reported \cite{Denis_NCEL_PRD} a flux averaged differential cross section (Fig.~\ref{fig:1}) in terms of momentum transferred squared to the nucleon, $Q^{2}_{QE}$. $\textit{Note:}$ In MiniBooNE  $Q^{2}_{QE}$ is the total kinetic energy of the outgoing nucleons in the interaction, assuming the target nucleon is at rest.
$$
 Q^{2}_{QE}=2M_{N}T=2M_{N}\sum_{i}T_{i}
$$
where $M_{N}$ is the nucleon mass and T is the sum of the kinetic energies of the final state nucleons. This makes the MiniBooNE NCE cross section measurement less sensitive to final state interactions (FSI) as the total energy measured in the detector is the same due to energy conservation, even if the energy of the neutrino may be divided among many outgoing nucleons. It should be noted that this scattering is off both bound nucleons in  carbon and free nucleons in the hydrogen atom of the target mineral oil ($CH_{2}$).

\begin{figure}
\includegraphics[scale=0.42]{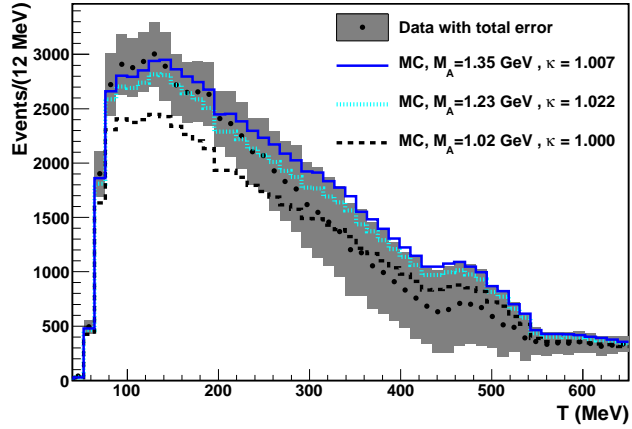}
\caption{\textit{(color online)} NCE reconstructed nucleon kinetic energy distribution with MC with different $M_{A}$ values of 1.35, 1.23, and 1.02 GeV. The $\chi^{2}$ values are 27.1, 29.2, and 41.3 for 49 degrees of freedom (DOF), respectively. The distributions are absolutely normalized.}
\label{fig:2}
\end{figure}


        A $\chi^{2}$ goodness of fit test was performed using the reconstructed NCE energy spectrum to find the set of $M_{A}$ and $\kappa$ (Pauli blocking scaling parameter) that best matches data.  Fig.~\ref{fig:2} shows the reconstructed energy spectrum for data and monte carlo(MC) prediction for different values of $M_{A}$ and $\kappa$--measured from MiniBooNE CCQE data \cite{MB_CCQE} and values prior to MiniBooNE.  Assuming $\Delta s$=0, the MiniBooNE NCE sample yields:
$$
M_{A}=1.39\pm 0.11 \textrm{GeV}
$$
with $\chi^{2}_{min}/DOF$=26.9/50.

\begin{figure}
\centering
\includegraphics[scale=0.42]{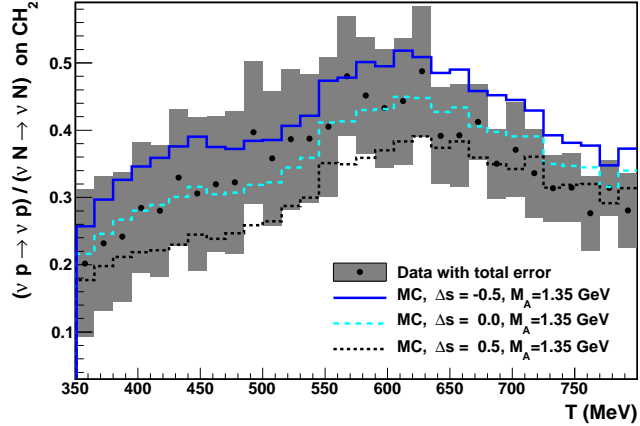}
\caption{\textit{(color online)} The ratio $\nu p\rightarrow\nu p$ to $\nu N\rightarrow\nu N$ as a function of  reconstructed energy for data and MC with $\Delta s$ values as shown.}
\label{fig:3}
\end{figure}

      Even though the ratio $\nu p\rightarrow\nu p$ to $\nu n\rightarrow\nu n$ (n being neutron) is more sensitive to $\Delta s$ \cite{wtf}, in MiniBooNE a neutron can only be detected if it has a further strong interaction with a proton, which at low energies is difficult to distinguish from single proton events. Hence a sample of single protons above 350 MeV (the Cerenkov threshold for protons in MiniBooNE) was used and the ratio of $\nu p\rightarrow\nu p$ to $\nu N\rightarrow\nu N$ (N being nucleon) as a function of reconstructed nucleon energy from 350 MeV to 800 MeV was studied to measure $\Delta s$ (Fig.~\ref{fig:3}). Additionally looking at such a ratio reduces the effect of FSIs and also some systematic errors.
Assuming  $M_{A}$= $1.39\pm 0.11$ GeV the $\chi^{2}$ tests of $\Delta s$ to MiniBooNE measured $\nu p\rightarrow\nu p /\nu N\rightarrow\nu N$ gives:
$$
\Delta s=0.08\pm 0.26
$$
with $\chi^{2}_{min}/DOF$=34.7/29. This is consistent with the BNL E734 measurement \cite{BNL_E734}.

\begin{figure}
\begin{minipage}[t]{0.5\linewidth}
\centering
\includegraphics[scale=0.36]{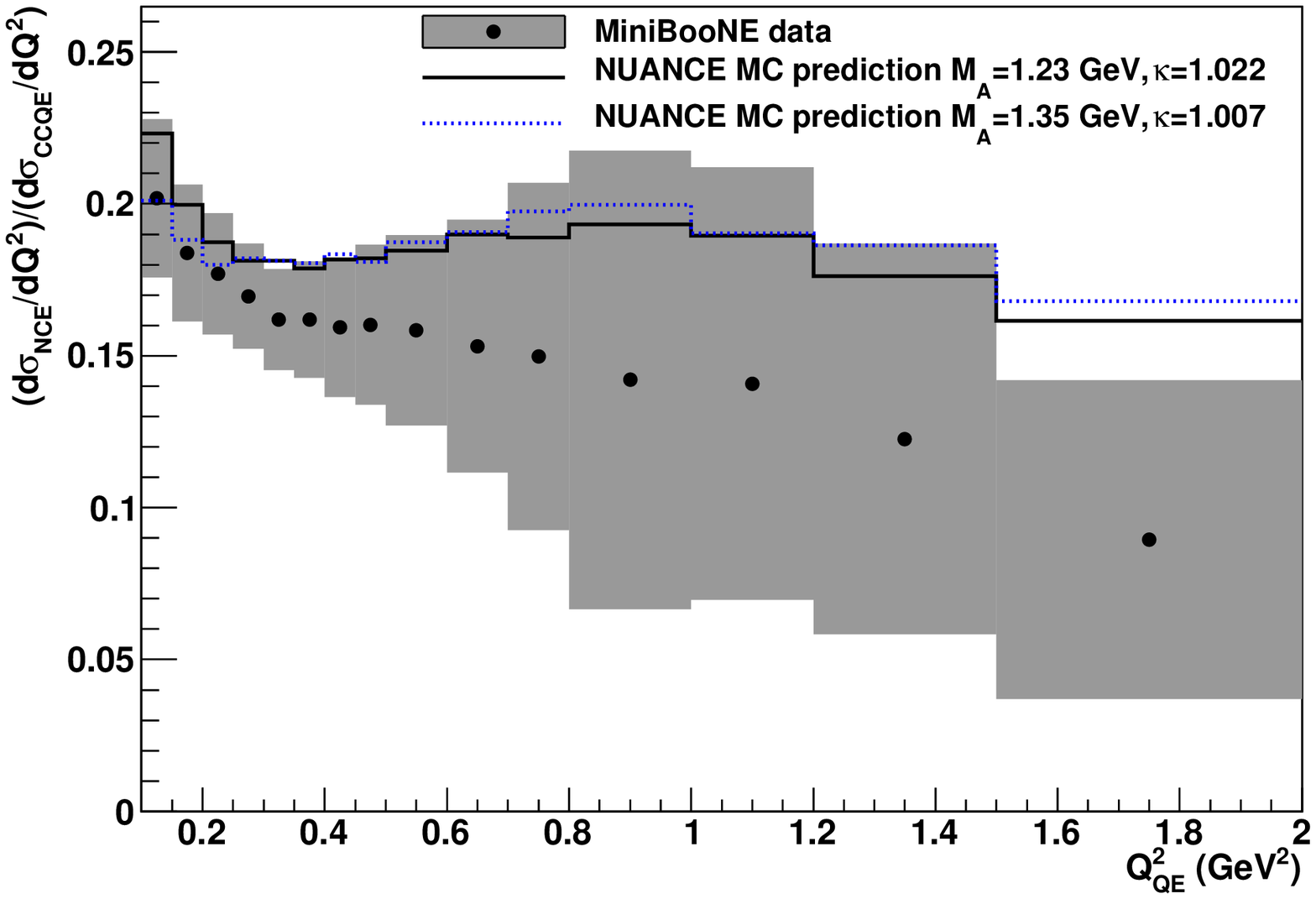}
\caption{\textit{(color online)} $\it{Left:}$ MiniBooNE NCE/CQE cross section ratio as a a function of $Q^2_{QE}$ Also shown is MC prediction with different $M_A$ and $\kappa$ values. $\it{Right:}$ MiniBooNE NCE-like/CCQE-like cross section ratio where the model dependent backgrounds are accounted as signal.}
\label{fig:4}
\end{minipage}
\hspace{0.5cm}
\begin{minipage}[t]{0.5\linewidth}
\centering
\includegraphics[scale=0.36]{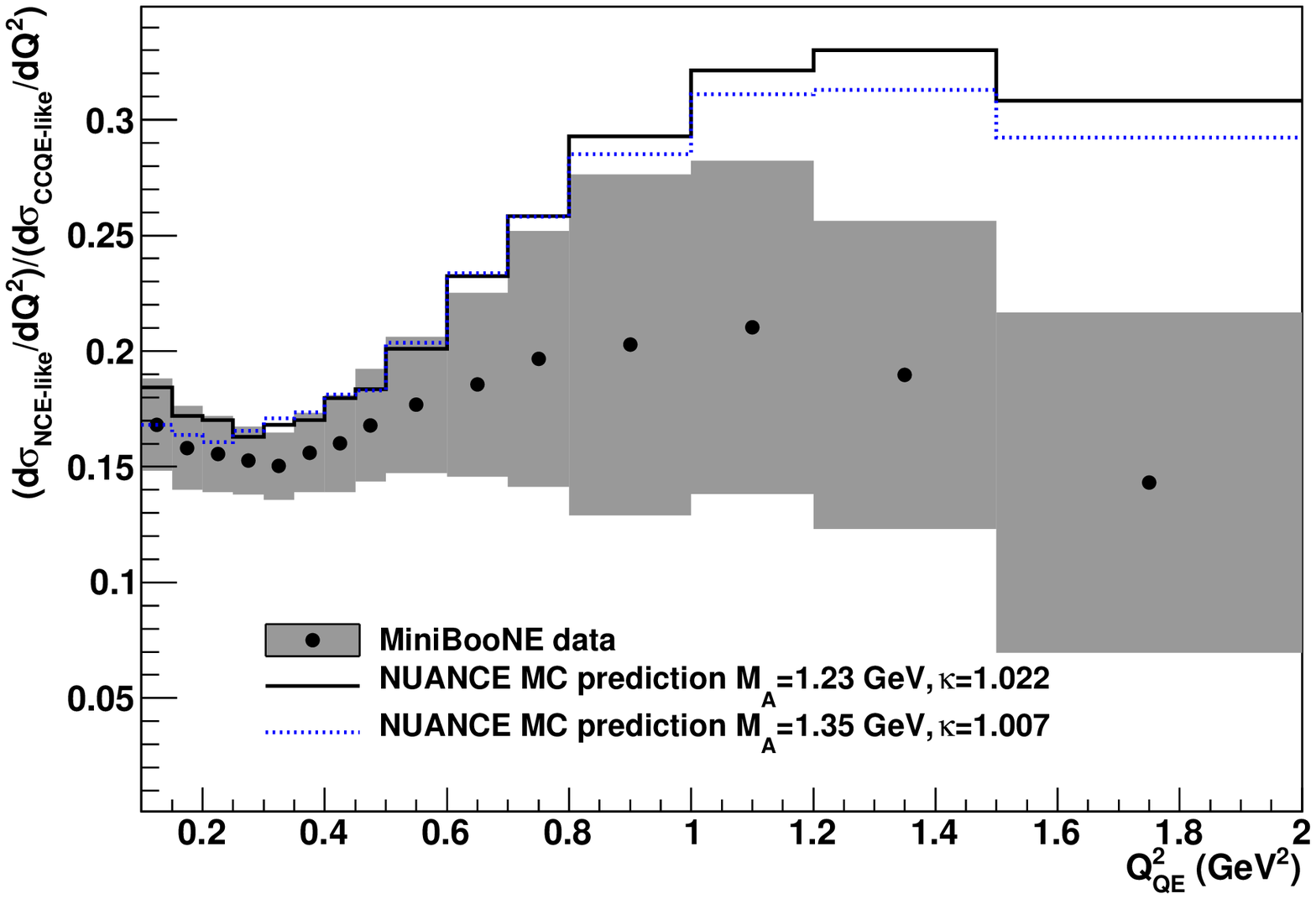}
\end{minipage}
\end{figure}

 To facilitate a comparison between the MiniBooNE CCQE differential cross section \cite{MB_CCQE} and the NCE measurement, a NCE/CCQE ratio as a function of $Q^{2}_{QE}$ was extracted. MiniBooNE measures the two cross sections differently--in case of CCQE, $Q^2_{QE}$ is defined from outgoing muon kinematics and assuming a stationary neutron. The ratio measurement also reduces the flux uncertainties.
Fig.~\ref{fig:4}(left) shows the measured ratio with the MC prediction for the same. This is the first time such a ratio measurement has been made.

Fig.~\ref{fig:4}(right) shows the NCE-like/CCQE-like ratio which is a more model independent comparison made by not subtracting the non-elastic backgrounds from the signal in both measurements in the ratio. Also shown is the MC prediction with different values of $M_{A}$ and $\kappa$.

\section{new results from anti-neutrino mode}
Before looking at $\bar\nu$ NCE interactions, it is important to measure a significant background in the $\bar\nu$ mode beam, the $\nu$'s in the $\bar\nu$ beam--the so called wrong sign (WS) background. The $\bar\nu$ beam  has a higher $\nu$ background component, as compared to the corresponding $\bar\nu$ background in  $\nu$ mode. The WS background in MiniBooNE has been measured using three independent techniques \cite{Joe_WS1}.

\begin{figure}[t]
  \includegraphics[height=.33\textheight]{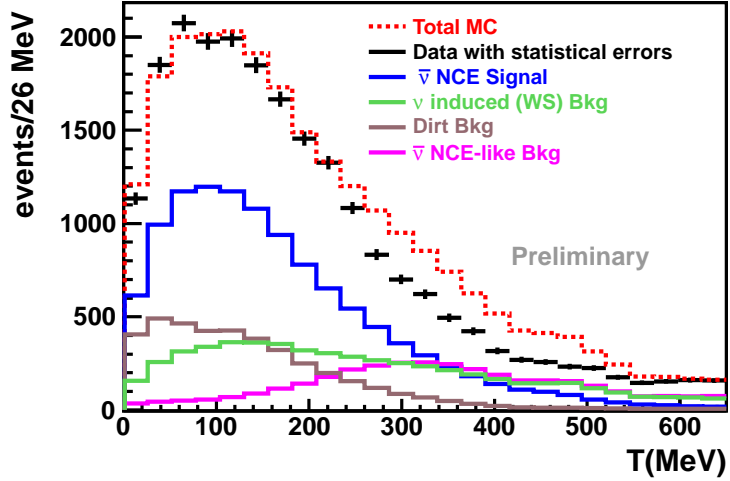}
  \caption{\textit{(color online)} $\bar\nu$ NCE data with statistical error (black), NCE signal MC(blue), $\nu$ induced WS background MC(green), $\bar\nu$ NCE-like background (purple), dirt MC(brown) and total MC (red).}
\label{fig:5}
\end{figure}

After applying the $\bar\nu$ NCE selection criteria, we have about 21,500 events corresponding to $4.88\times10^{20}$ POT. Fig. \ref{fig:5} shows the measured $\bar\nu$ NCE energy spectrum data with statistical errors. Also shown is the $\bar\nu$ NCE MC signal prediction (47\%) and the major $\it{backgrounds}$--dirt events (17\%), $\nu$ induced WS events (22\%),  and  $\bar\nu$ NCE-like background (14\%).

The low energy dirt background is due to interactions in the dirt surrounding the detector. This background has  been measured using MiniBooNE data (see Appendix A in Ref \cite{Denis_NCEL_PRD} for details). The $\nu$ induced WS background is measured as mentioned earlier.  $\bar\nu$ NCE-like background are NC pion events with pion absorption which mimics the $\bar\nu$ NCE signal in MiniBooNE (hence referred to as $\bar\nu$ NCE-like). We estimate  $\bar\nu$ NCE-like background using our MC prediction.

\begin{figure}[t]
\includegraphics[height=.33\textheight]{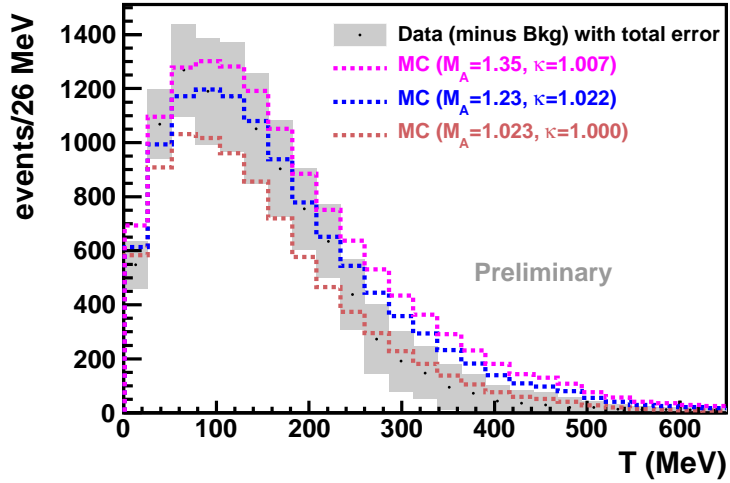}
\caption{\textit{\textit{(color online)}} $\bar\nu$ NCE reconstructed energy distribution with error(systematic and statistical) with MC predictions for different $M_A$ and $\kappa$}
\label{fig:6}

\end{figure}

Fig.~\ref{fig:6} shows the reconstructed $\bar\nu$ NCE energy spectrum with the total error(systematic + statistical) after all backgrounds have been subtracted. Also shown for comparison are MC predictions with different values of model parameters: $M_A$ and $\kappa$. 

\section{summary}
The MiniBooNE experiment has collected the largest sample to date of NCE interactions both in $\nu$ and $\bar\nu$ mode offering a unique opportunity for comparison. The  results from the $\nu$ mode \cite{Denis_NCEL_PRD, Denis_thesis} have been previously reported. The $\bar\nu$ mode analysis is in progress with plans to report:  the flux averaged differential cross section, measurements of $M_{A}$, $\Delta s$ and ratio measurements--comparing the $\bar\nu$ NCE with CCQE in both $\nu$ and $\bar\nu$ mode.


\begin{theacknowledgments}
Our thanks to the organizers of NuInt'11 for the opportunity to present these results.
\end{theacknowledgments}



\bibliographystyle{aipproc} 

\bibliography{refer}

\end{document}